\documentclass{elsart}                                                
 \usepackage{amsmath,amssymb,epsfig}
 \def\be{\begin{eqnarray}}
 \def\ee{\end{eqnarray}}

 \newcommand{\bk}{{\bf k}}
 
 \newcommand{\br}{{\bf r}}

\begin{document}
 \begin{frontmatter}

 \title{Pairing in nuclear systems with 
 effective Gogny and $V_{\rm low~k}$ interactions}
 \author{A. Sedrakian$\hspace{1mm}^{a}$},
 \author{T.T.S. Kuo$\hspace{1mm}^{b}$},
 \author{H. M\"uther$\hspace{1mm}^{a}$} and 
 \author{P. Schuck$\hspace{1mm}^{c}$} 
 \address{$^{a}$ Institut f\"ur Theoretische Physik, 
        Universit\"at T\"ubingen, 
        D-72076 T\"ubingen, Germany\\
 $^{b}$Department of Physics and Astronomy, State University of
 New York,\\
 Stony Brook, NY 11794-3800, U. S. A.\\
 $^{c}$Institut de Physique Nucleaire, 
        IN2P3-CNRS, Universit\'{e} Paris-Sud, F-91406 Orsay Cedex, 
         France\\}

 \begin{abstract}
 The pairing properties of nuclear systems are a sensitive probe of the
 effective nucleon-nucleon interactions. We compare the  $^1S_0$
 pairing gaps in nuclear and neutron matter derived from 
 the phenomenological Gogny interaction and  a renormalization group 
 motivated low-momentum  $V_{\rm low~k}$ 
 interaction extracted from realistic interactions. 
 We find that the pairing gaps predicted by these interactions 
 are in an excellent agreement in a wide range of sub-nuclear
 densities. The close agreement between the predictions of the
 effective forces remains intact in the case where the
 single particle spectra in neutron and nuclear matter are
 renormalized by nuclear interactions.


 \end{abstract}
 \end{frontmatter}

 The effective nuclear forces are the key ingredients in the nuclear
 structure studies of finite nuclei within the Hartree-Fock-Bogoliubov
 and related mean field theories. Well-known examples are the zero-range
 Skyrme \cite{SKYRME} and finite-range Gogny \cite{GOGNY1,GOGNY2}
 forces that have extensively been used in the large 
 scale numerical calculations of finite nuclei over several decades. 
 The effective forces are commonly adjusted to the bulk properties of
 nuclear systems after a mean-field variational minimization of the
 ground state energy of a collection of nuclei within the density
 dependent Hartree-Fock or Hartree-Fock-Bogoliubov theories.

 A distinctive feature of the {\it finite range} Gogny forces, 
 which will be discussed below, is that these were adjusted by 
 minimizing the Hartree-Fock-Bogoliubov energy functional that 
 is a function of the pairing fields~\cite{GOGNY1,GOGNY2}. Thus, the
 pairing correlations in the system, seen experimentally 
 in the  odd-even staggering 
 effects caused by the pairing in the isospin $T=1$ states, are
 encapsulated in the effective force. 
 Since, in general, the effective forces are subject to simple
 parameterizations and 
 have the advantage of reducing the numerical cost of the
 extensive nuclear structure calculations, it remains an important task
 to scrutinize the reliability of effective forces in different
 contexts, in particular their relation to the interactions derived from
 the underlying microscopic theories.

 During the recent years much effort went into formulations of 
 nuclear interactions in terms of effective field theories
\cite{weinberg,kaplan98,lepage,shankar,EFT,haxton,epelbaum1,epelbaum2}.
 The main
 goal of these theories is the separation of the long-range component
 of the nuclear forces, which is dominated by the pion exchange and is
 well under control, from the intermediate and 
 short range components, which are dominated by correlated pion and 
 heavy meson exchanges that are poorly known. 

 A line of approach, developed by the Stony-Brook
 group, applies the renormalization group arguments to the
 Lipmann-Schwinger (LS) equation to eliminate the high-momentum 
 modes of a phase-shift equivalent potential 
 $V_{NN}$ which serves as a driving term in the LS equation
 \cite{bogner01,bogschw01,bogner02,schwenk02,coraggio02a,coraggio02b,bogner03}.
  Note that low momentum nucleon-nucleon 
  (NN) interactions were also derived by 
  the Bochum-J\"ulich group by applying a method of unitary transformations 
  to full (un-truncated) meson exchange interactions derived from
  chiral Lagrangians \cite{epelbaum1,epelbaum2}.
  The Stony-Brook low-momentum NN interaction $V_{\rm low~ k}$ is obtained by 
  integrating out the high momentum components
  of $V_{NN}$ beyond a scale $\Lambda$.   The following LS
  equations for the scattering amplitudes 
  with driving terms $V_{NN}$ and $V_{\rm low~k}$ are considered:
 \begin{equation}
   T(k',k,k^2)  
 = V_{NN}(k',k)   
  + \int _0 ^{\infty} q^2 dq  V_{NN}(k',q) 
  \frac{1}{k^2-q^2 +i0^+ } T(q,k,k^2 ) ,
 \end{equation}
 \begin{equation}
   T_{\rm low~k }(p',p,p^2)  
 = V_{\rm low~k }(p',p)   
  + \int _0 ^{\Lambda} q^2 dq  V_{\rm low~k }(p',q) 
  \frac{1}{p^2-q^2 +i0^+ } T_{\rm low~k} (q,p,p^2).
 \end{equation}
 Note that the intermediate state momentum $q$ is integrated from
 0 to $\infty$ and 0 to $\Lambda$ in the first and second equation
 respectively. The equivalence of the $T$-matrices derived from the 
 LS equation above is required:
  $T(p',p,p^2 ) = T_{\rm low~k }(p',p, p^2 ) ;~( p',p) \leq \Lambda$.
 As described in Ref. \cite{bogner02}, the $V_{\rm low~k}$ interactions
 are then derived by the Andreozzi-Lee-Suzuki method \cite{andre96}.

 The $V_{\rm low~k}$ so derived reproduces the empirical deuteron
 binding energy, NN scattering phase shifts up to $E_{\rm lab}=
 2\hbar^2\Lambda^2/M$.
 Experiments give us information about phase shifts only up to
 $E_{\rm lab}\sim$ 350 MeV. Thus an appropriate choice for $\Lambda$
 is $\sim 2$ fm$^{-1}$. Beyond this momentum, 
  $V_{NN}$ is model dependent and lacks physical ground. 
 An interesting
 feature of the resulting effective interaction  $V_{\rm low~k}$ is
 that it is  largely independent of the underlying
 microscopic force that was used as a driving term in the LS equation
 \cite{bogner03}.
 This property clearly reflects the fact that the low-momentum part of
 the microscopic interactions is well constrained by the experimental
 data, i.e., they equally well reproduce the binding energy of the
 deuteron and are phase-shift equivalent~\cite{bogner02,bogner03}.
 The initial applications of the $V_{\rm low~k}$ interaction are
 in the shell model calculations
 \cite{bogner02,coraggio02a,coraggio02b} and in various treatments of 
 the equation of state and the pairing in nuclear systems 
 \cite{schwenk02,kuckei}.
 Clearly, the renormalization group based 
 decimation procedure, adopted to derive the 
 effective $V_{\rm low~k}$ interaction, bridges the gap
 between the effective and microscopic interactions in a controlled
 manner, thus it provides  a good starting point for mean-field
 calculations. 

 The purpose of this {\it Letter} is to compare the predictions of the Gogny and 
 $V_{\rm low~k}$ interactions for the pairing in nuclear systems in the 
 $^1S_0$ interaction channel.  We will be concerned with
 infinite, zero-temperature matter, parameterized in terms of the
 Fermi-momentum $k_F$ and will mainly focus on effects of the force on 
 pairing. We shall briefly comment on the renormalization of the single
 particle spectrum in the mean-field approximation, but
 leave aside the issues of the vertex and propagator
 renormalizations beyond the mean field. Some aspects of the
 pairing, which are complementary to this study, are explored in Refs.
 \cite{schwenk02,kuckei} and \cite{GOGNY_PAIRING1,GOGNY_PAIRING2}
 using the $V_{\rm low~k}$ and Gogny interactions, respectively.
 Our work, in part, is motivated by the observation that the Gogny
 interactions predict pairing properties that are surprisingly close
 to those derived from the bare realistic (phase shift equivalent)
 nuclear interactions \cite{GOGNY_PAIRING1}.

 The effective Gogny interactions are of the generic form 
 \begin{eqnarray} \label{GOGNY}
 V(\br_1-\br_2)  &=&  t_0 (1 + x_0 P_{\sigma})\delta(\br_1-\br_2) \rho^{d}
 \nonumber\\
 &&\hspace{-1.cm}+\sum_{m=1}^2 [W_{m} + B_{m} P_{\sigma} -
 H_{m} P_{\tau} - M_{m} P_{\sigma} P_{\tau}]~ {\rm exp}
 \left(- \frac{\vert \br_1-\br_2\vert^2}{\mu_m^2}\right),
 \end{eqnarray}
 i.e. they contain two Gaussian terms which reflect  the finite-range
 of the interaction, and a contact density-dependent term
 responsible for short range correlations. Here
 $\rho$ is the density and $P_{\sigma}, P_{\tau}$ are the 
 spin and isospin exchange operators. For completeness we reproduce 
 the values of the  parameters according to the
 D1S parameterization in Table 1 \cite{GOGNY2}. 
 (Compared to the original D1 parameterization \cite{GOGNY1}
 the D1S force 
 gives a lower surface tension  and, at the same time, 
 a smaller even-odd staggering which
 is closer to the experimental values.) 
 \begin{table}[b]
 \begin{center}
 \begin{tabular}{cccccc}
 \hline
 $m$ & $\mu_m$ & $W_m$ & $B_m$ &  $H_m$ &  $M_m$ \\
 \hline 
 1  &      0.7  &  -1720.3    &   1300.0    &  -1813.53  & 1397.6\\
 2  &      1.2  &   103.639   &   -163.483  &    162.812 & -223.934\\
 \hline
 \end{tabular}
 \caption{\label{tab1} The parameters of the D1S Gogny interaction \cite{GOGNY2}.
 The ranges of the interaction $\mu_m$ are in fm while the remainder
 coefficients are in MeV. The values of the parameters of the contact
 term are $t_0 = 1390$ MeV fm$^4$,  $x_0 = 1$, and $d = 1/3$.
 }
 \end{center}
 \end{table}

 The $V_{\rm low~k}$ interaction we shall employ is based on the
 Nijmegen  93 potential with a momentum cut-off $\Lambda = 2.5$
 fm $^{-1}$, which corresponds to a distance scale $\Lambda^{-1}\sim$
 0.4 fm. The choice of the underlying
 realistic potential is not important, as the effective interactions
 derived from various realistic interactions are practically identical
 (c.f. ~\cite{bogner03}).

 The gap in the quasiparticle spectrum of infinite nuclear systems is
 governed by the Bardeen-Cooper-Schrieffer (BCS) integral equation for the 
 gap function 
 \begin{equation}
 \Delta(k)= -\frac{1}{2} \int_0^\Lambda dk'\,k'^2\, V(k,k') \frac{\Delta(k')}
 {\sqrt{\left(\varepsilon_{k'} - \varepsilon_F\right)^2 + \Delta(k')^2}},
 \label{eq:gap}
 \end{equation}
 where $V(k,k')$ is the momentum space effective pairing interaction,
 $\varepsilon_k$ is the quasiparticle spectrum, $\varepsilon_F$ is the
 Fermi-energy, and $\Lambda$ is momentum space cut-off. The
 high-momentum behaviour of the pairing interaction, for the case of 
 Gogny force, is controlled by the finite range of the Gaussians;
 hence we can take safely the limit  $\Lambda \to \infty$ in 
 Eq. (\ref{eq:gap}). The $V_{\rm low~k}$ interactions have a sharp
 cut-off at high-momenta, and it is natural to identify 
 $\Lambda$ with the cut-off in the interaction. 
 Note that when transformed in the momentum space
 the Gogny interaction depends on the momentum transfer in the process,
 and hence on the angle between the relative incoming and outgoing
 momenta of the particles. It is then suitable to average the matrix
 elements over the angle between the vectors $\bk$ and $\bk'$. The
 angle averaged $^1S_0$-wave pairing interaction  can be 
 written as~\cite{GOGNY_PAIRING1}
 \be \label{GOGNY3}
 V(k,k') &=& \frac{1}{\sqrt{\pi}kk'} \sum_{m=1}^2 C_m{\rm exp}
 \left[-\frac{\mu_m^2}{4}(k^2+k'^2)\right]
 {\rm sinh}\left(\frac{\mu_m^2kk'}{2}\right),
 \ee 
 where $C_m \equiv \mu_m(W_m-B_m-H_m+M_m)$. Note that only the 
 density independent  part of the Gogny interaction contributes to
 the pairing in the $^1S_0$-channel. The angle averaged pairing
 interaction thus contains two Gaussian describing long range 
 attraction and a short range repulsion (the terms $\propto C_1$
 and $C_2$, respectively).

 Next we need to specify the single particle spectrum, $\varepsilon_k$,
 in the gap equation (\ref{eq:gap}). 
 We shall employ two approximations. First, the single particle 
 spectrum in a non-interacting limit will be used to  
 understand the correlations between the pairing gap and the 
 pairing force. Second, the single particle spectrum will be
 renormalized within the mean-field approximation. We shall use the 
 Hartree-Fock single particle spectra for the Gogny interaction, which
 are derived below.
 For the $V_{\rm low~k}$ interaction we shall employ
 the Brueckner-Hartree-Fock scheme (see Ref. \cite{kuckei} and
 references therein.) 
 It should be kept in mind that when the effective interactions depend on  
 energy, e. g. when one is dealing with time-retarded interactions, 
 the wave-function renormalization differs from unity 
 and tends to counter-act the reduction of the mass caused by the
 momentum dependent self-energies. Thus, the net effect of medium
 renormalization of particle mass could be an overestimate.

 To define the single-particle spectrum, it is useful to start with 
 the expression for the ground state energy (at zero temperature) 
 in the mean-field approximation 
 \be\label{GR_ENERGY} 
 E = \sum_{i} \frac{\hbar^2k_i^{2}}{2 m}
 n_{i} + {\frac{1}{2}} \sum_{i j} \langle i j \mid  V
 \mid i j - j i \rangle n_{i} n_{j},
 \ee
 where the first term is the kinetic energy, 
 the second term is the potential energy of the mean-field
 interaction; the indices refer to the nucleonic states,
 $n_i$ are their occupation probabilities.
 In the case where the interaction is density dependent the single
 particle potential is given by the functional derivative of the second 
 term $(E_{\rm int})$ in Eq. (\ref{GR_ENERGY}), i.e. 
 $U_{i} \equiv  \delta E_{\rm int}/\delta n_{i}$, or explicitely
 \be\label{U} 
 U_{\tau i}  =  \sum_{j}  \langle i j \mid  V \mid i j - j i\rangle n_{j} +
 {\frac {1}{2\Omega}} \sum_{j l} \langle j l \mid {\frac
 {\partial  V}{\partial\rho_{\tau}}} \mid j l - l j\rangle n_{j} n_{l},
 \ee
 where $\tau (= n,p)$ is the isospin index, $\Omega$ is the volume. 
 Equation (\ref{U}) should be solved 
 self-consistently with the normalization condition for the 
 total density $\sum_{i} n_{i}  =\rho$. 

 The matrix elements defining the single-particle potential 
 are evaluated in the mean-field approximation 
 using plane waves for the nucleon states. 
 We find  
 \be \label{UPOT}
 U_{\tau}(k) &=&  \frac{t_0}{4}(2+x_0)(2+d) \rho^{d+1}
 - \frac{t_0}{4}(1+2x_0)
 \left[{2x_{\tau}}+d (x_p^2+x_n^2)\right] \rho^{d+1}\nonumber\\
 & + &
 \frac{1}{2}\sum_{m=1}^2 \rho  F_m(0)\left[
 \left(2W_m+{B_m}\right) -\left(2H_m+{M_m}\right)x_{\tau}
 \right]\nonumber\\
 &+&\sum_{m=1}^2 \sum_{\bk'}   F_m(\bk-\bk')
 \Bigl[\left(H_m+2M_m\right)\left[n_p(k')+n_n(k')\right]
 -\left(W_m+2B_m\right)n_{\tau}(k')
 \Bigr],\nonumber\\
 \ee
 where $x_{\tau} = \rho_{\tau}/\rho$ are the relative concentrations of the 
 neutrons and protons, $n_{\tau}(k) = \theta(k_{F\tau}-k)$ are their
 occupation probabilities, and we defined a momentum dependent form-factor as
 \be F_m(\bk-\bk') & = &
 \pi^{3/2} \mu_m^{3} e^{-\mu_m^{2} \vert\bk - \bk'\vert ^{2}/4}.
 \ee
 The first two terms in Eq. (\ref{UPOT}) correspond to the contributions
 of the contact interaction, where the terms proportional to $d$
 originate from the rearrangement interaction. The 
 third and the fourth terms are  the direct and the 
 exchange contributions of the finite range part of interaction. 
 In the zero-temperature limit of interest
 the phase space integrals over the Fock 
 term in Eq. (\ref{UPOT})  can be done analytically, 
 \be\label{UFOCK} 
 U_{\tau}^{\rm Fock}(k) &=& 
 -\frac{1}{2}\sum_{m=1}^2\left[W_m+2(B_m-M_m)-H_m\right]\nonumber\\
 &&\Biggl\{\frac{2}{\sqrt{\pi}\mu_mk}\left[
 {\rm exp}\left(-\frac{\mu_m^2}{4}(k+k_{Fq})^2\right)
 -{\rm exp}\left(-\frac{\mu_m^2}{4}(k-k_{Fq})^2\right)
 \right]\nonumber\\
 &&\hspace{4cm}+{\rm Erf}\left[\frac{\mu_m}{2}(k+k_{Fq})\right]
 -{\rm Erf}\left[\frac{\mu_m}{2}(k-k_{Fq})\right]\Biggr\},
 \nonumber\\
 \ee
 and the net single particle energy $U_{\tau} = U_{\tau}^{\rm
 Hartree}+U_{\tau}^{\rm
 Fock}$, where $U_{\tau}^{\rm Hartree}$ corresponds to the first three terms 
 in Eq. (\ref{UPOT}), becomes an analytical function of the Fermi-momenta
 of neutrons ($k_{Fn}$) and protons ($k_{Fp}$). The single particle
 spectrum is then defined as 
 $
 \varepsilon_{\tau}(k) = k^2/2m+ U_{\tau}(k).
 $
 Although the equations above are specified for arbitrary isospin 
 asymmetry, we shall further concentrate on the two special 
 cases $x_p=x_n$ (symmetric nuclear matter) and $x_p = 0$ (pure neutron
 matter).

 To understand the correlations between the effective pairing
 interactions and the pairing gap it is useful to fix the argument of
 the gap function on the left-hand-side of Eq. (\ref{eq:gap}) at
 the Fermi-momentum, i.e., 
 \begin{equation}
 \Delta(k_F)= 
 -\frac{1}{2} \int_0^\Lambda dk'\,k'^2\, V(k_F,k') \frac{\Delta(k')}
 {\sqrt{\left(\varepsilon_{k'} - \varepsilon_F\right)^2 + \Delta(k')^2}}.
 \label{eq:gap2}
 \end{equation}
 The kernel of the gap equation is a product of the momentum space
 matrix element and $V(k_F,k)$ which, as we shall see below, is a
 smooth function of momentum $k$ and the 
 anomalous propagator (the remainder multiplier in the kernel) 
 which is bell-shaped with the maximum $1/2$ for  
 $\varepsilon_k = \varepsilon_F$. Since the main contribution to the
 integral comes from the vicinity of the Fermi-surface the differences
 in the pairing gaps reflect the differences in the 
 effective forces in the vicinity of the Fermi-surface. This
 observation  motivates an approximation where we replace
 $V(k_F,k)$ by $V(k_F,k_F)$. Such an approximation permits to solve 
 Eq. (\ref{eq:gap2}) analytically and the result is the well know BCS 
 weak-coupling formula. 
 \begin{table}[t]
 \begin{center}
 \begin{tabular}{ccccccc}
 \hline
   $k_F$   &  $\Delta_{\rm Gogny}$  &  $\Delta_{\rm Vlow~k}$ &  $R_{\Delta}$ &
 $V_{\rm Gogny}$  &  $V_{\rm low~k}$ &  $R_{V}$\\
 \hline
 0.1  & 0.05  &  0.10  & 0.552& -39.5 & -49.1 & 0.80 \\
 0.2  & 0.35  &  0.46  & 0.758& -38.4 & -46.5 & 0.83 \\
 0.3  & 0.83  &  0.98  & 0.852& -36.6 & -42.9 & 0.85 \\
 0.4  & 1.42  &  1.57  & 0.907& -34.3 & -38.9 & 0.88 \\
 0.5  & 2.03  &  2.15  & 0.946& -31.6 & -34.8 & 0.91 \\
 0.6  & 2.60  &  2.66  & 0.978& -28.6 & -30.8 & 0.93 \\
 0.7  & 3.09  &  3.04  & 1.006& -25.4 & -26.9 & 0.94 \\
 0.8  & 3.38  &  3.26  & 1.037& -22.2 & -23.2 & 0.96 \\
 0.9  & 3.52  &  3.27  & 1.076& -19.2 & -19.7 & 0.97 \\
 1.0  & 3.47  &  3.06  & 1.131& -16.3 & -16.5 & 0.99 \\
 1.1  & 3.22  &  2.64  & 1.220& -13.7 & -13.5 & 1.01 \\
 1.2  & 2.81  &  2.02  & 1.391& -11.3 & -10.7 & 1.06 \\
 1.3  & 2.26  &  1.28  & 1.766& -9.3  & -8.3  & 1.12 \\
 1.4  & 1.66  &  0.56  & 2.964& -7.6  & -5.9  & 1.28 \\
 \hline
 \end{tabular}
 \caption{\label{tab2} The pairing gaps for
 different Fermi-momenta $k_F$ computed with the Gogny (second column)
 and the $V_{\rm low~k}$ interactions (third column). The fourth column
 shows the ratio, $R_{\Delta}$, of the gap values given 
 in the second and third columns to make the variations 
 with the effective interaction explicite. The fifth and sixth columns
 show the diagonal elements of the interactions $V_{\rm
 Gogny}(k_F,k_F)$  and $V_{\rm low~k}(k_F,k_F)$ and the last 
 column - their ratio, $R_{V}$.
 }
 \end{center}
 \end{table}
 Table 2 lists the values of the gaps calculated 
 numerically (without approximations) using free single-particle
 spectrum  along with the diagonal elements of the effective
 interactions. A clear correlation is seen between the ratios of the 
 gaps derived from Gogny and $V_{\rm low~k}$ interactions and corresponding
 diagonal matrix elements (these ratios are related in a highly
 non-linear manner).
 It is remarkable that the predictions of the Gogny and  
 $V_{\rm low~k}$ interactions differ by no more than 
 10\% in a wide range of subnuclear densities ($0.4 \le k_F\le 1.0$ 
 fm$^{-1}$) which are relevant for the description of pairing phenomena in 
 finite nuclei. It is also seen that the pairing gaps are a sensitive
 probe of the interactions (as can be expected from the BCS weak 
 coupling result): a  20\% deviation in the forces results in a 
 factor of two deviation in the pairing gaps. Finally, it is worthwhile
 to note that, there is no such correlation for the realistic {\it bare}
 interactions, for the diagonal elements of, e.g., the Paris or the Reid
 interactions are repulsive.

 \begin{figure}
 \begin{center}
 \epsfig{figure=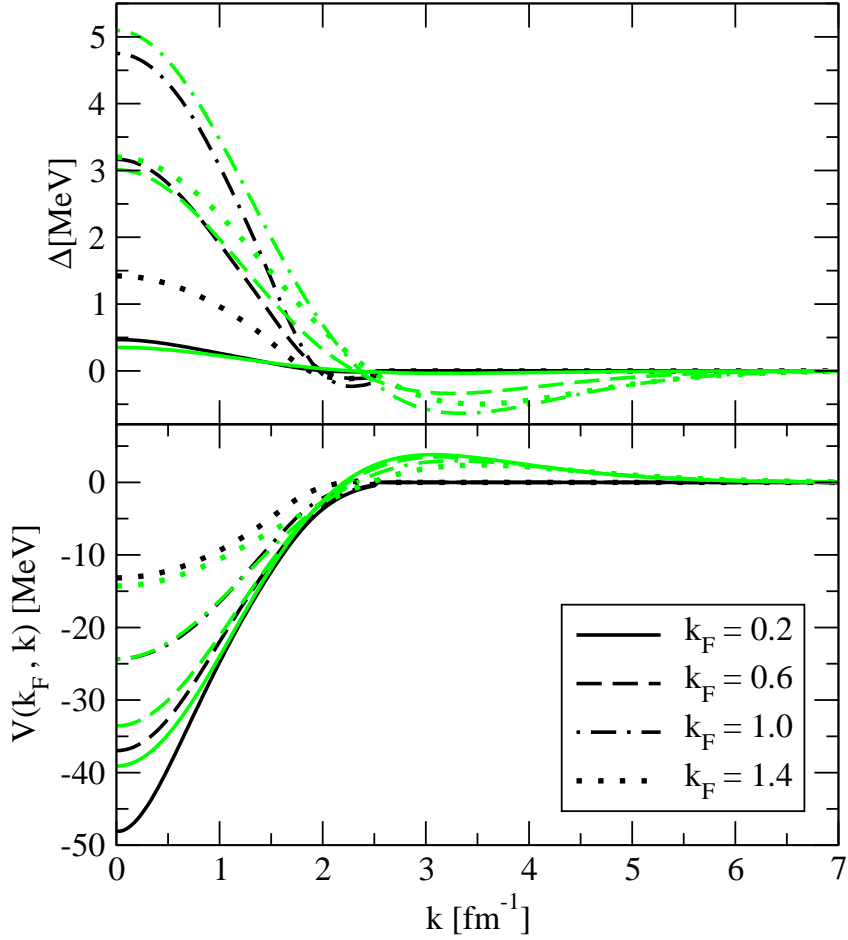,width=16cm,angle=270}
 \end{center}
 \caption{The pairing gaps in the $^1S_0$ channel and the corresponding
 pairing potentials $V(k_F,k)$ as functions of the momentum $k$ for several 
 fixed Fermi-momenta $k_F$. 
 The black and grey (green) lines refer to the $V_{\rm low~k}$ 
 and the  Gogny interaction respectively.
 The solid, dashed, dashed-dotted and dotted lines correspond to the 
 values of $k_F$ equal 0.2, 0.6, 1.0 and 1.4 fm$^{-1}$.}
 \label{fig1}
 \end{figure}

 The correlations between the momentum dependent pairing gaps
 $\Delta(k)$ calculated  numerically (without approximations) with 
 the momentum space matrix elements $V(k_F,k)$, which would correspond 
 to a pairing gap derived from Eq. (\ref{eq:gap2}), can be observed
 in Fig. 1. Quite generally, the momentum dependence of the pairing gap
 reflects the momentum dependence of the pairing potential. This
 relation becomes explicit if one approximates the pairing force by 
 a separable potential. Writing, schematically,  $V(k,k')=g(k)g(k')$ 
 and inserting this form in the gap equation, one finds that the solutions
 are of the form $\Delta(k)= C g(k)$ where $C$ is a constant.

 The $V_{\rm low~k}$ interaction is attractive below the cut-off scale 
 $\Lambda$ for the Fermi-momenta of interest and vanishes above it; 
 therefore there is always an associated non-zero solution to the 
 gap equation which likewise vanishes above the cut-off scale. The high
 momentum tail of the Gogny interaction ($k\ge \Lambda$) is slightly repulsive 
 and the pairing gap changes its sign above this scale and vanishes at
 much higher momenta; clearly, these high-momentum components do not
 affect the values of the gaps at their Fermi-surface, since the
 kernel of the gap equation is sharply peaked at the Fermi-momentum. 
 As seen in Fig. 1 the differences between the predictions of the 
 effective forces for the gap functions are closely correlated with 
 their differences in the vicinity of respective Fermi-momenta.

 \begin{figure}[ht]
 \begin{center}
 \epsfig{figure=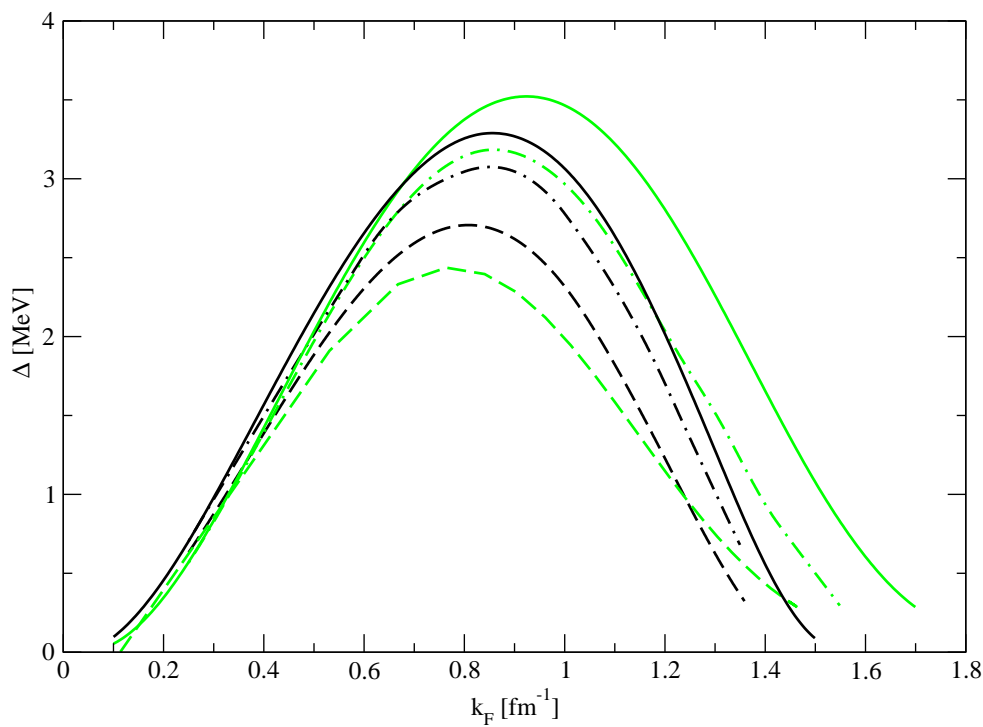,width=12cm,angle=270}
 \end{center}
 \caption{ The pairing gaps  in the $^1S_0$ channel as functions of
 Fermi-momentum (density) of the matter for two effective
 interactions.  The heavy and light lines refer to the $V_{\rm low~k}$ 
 and the  Gogny interaction respectively.
 The solid lines correspond to the non-interacting single-particle 
 spectrum, the dashed and  dashed-dotted lines - 
 to renormalized single particle spectrum in symmetric 
 nuclear matter  and neutron matter, respectively. 
 The single particle spectra are computed in the 
 Hartree-Fock theory for the Gogny interaction and the 
 Brueckner-Hartree-Fock theory for the  $V_{\rm low~k}$ interaction.}
 \label{fig2}
 \end{figure}

 Fig. 2 shows the density dependences of the pairing gaps in neutron 
 and nuclear matter ($k_F$ refers to the Fermi-momentum of nucleons 
 and neutrons respectively) and the effects of single particle 
 renormalization. The single particle spectra
 for the Gogny interactions were computed in the Hartree-Fock
 theory, as described above. For the $V_{\rm low~k}$ interactions we
 used the Brueckner-Hartree-Fock theory 
 with continuous choice of the single-particle spectrum. 
 If one adopts free single particle spectra for nucleons, the Gogny interaction
 predicts pairing gaps that are systematically larger 
 than those predicted by the $V_{\rm low~k}$ interaction for momenta larger 
 than about 1 fm$^{-1}$ and slightly smaller for momenta below this
 value; 
 this picture is consistent with the behaviour of the 
 potentials (see Fig.~1). The renormalization of the single particle
 spectra reduces the density of states at the Fermi-surface and,
 hence, the magnitude of the pairing gap. Independent of the chosen
 interaction this reduction is larger in the symmetrical nuclear
 matter than in the neutron matter, since in the former case the
 single particle spectra are steeper at the Fermi-surface mainly 
 due to  the tensor channel $^3S_1-^3D_1$ interaction.
 For neutron matter, where the
 renormalization of single particle spectrum is mild, there is a close
 agreement between the gaps computed with the two different interactions for
 $k_F\le 1$ fm$^{-1}$ and the deviations at larger densities are not
 dramatic. 
 For the symmetrical nuclear matter the deviations are larger,
 indicating that the differences in the single particle spectra are
 more important for the evaluation of the gap than the differences in
 the residual pairing interaction. 
 
 Extrapolations of the results above to finite nuclei depend on the
 extent the high-density region contributes to the average pairing gap. 
 If such a contribution is important, the pairing gaps predicted by the 
 $V_{\rm low~k}$ interaction must be systematically smaller that those 
 predicted by the Gogny interaction. A reduction of the pairing gaps 
 by a factor of two compared to the experimental values was observed with 
 realistic Argonne interaction in Sn isotopes \cite{VIGEZZI}, which 
 suggests that this might also be the case for the $V_{\rm low~k}$ interaction.

 To conclude, we observed a close agreement in the predictions of the
 pairing properties of nuclear systems by two effective interactions
 which have largely different origins - the Gogny phenomenological
 interaction, with parameters fitted  to reproduce nuclear properties
 in Hartree-Fock-Bogoliubov calculations, and the $V_{\rm low~k}$
 interaction, which is a renormalization group motivated  
 low-momentum reduction of the realistic interactions.We
 made explicit the correlations between the values of the pairing gaps
 and the diagonal and ``half on the Fermi-surface'' matrix elements of the 
 effective interactions.



 \end{document}